\documentclass[conference]{IEEEtran}
\makeatletter
\def\ps@headings{%
\def\@oddhead{\mbox{}\scriptsize\rightmark \hfill \thepage}%
\def\@evenhead{\mbox{}\scriptsize\leftmark \hfill \thepage}%
\def\@oddfoot{}%
\def\@evenfoot{}}
\makeatother
\usepackage[pdftex]{graphicx}
\usepackage[lined]{algorithm2e}
\usepackage{float}
\usepackage{epstopdf}
\newfloat{Algorithm}{t}{lop}

\ifCLASSINFOpdf
\else
\fi
\usepackage[cmex10]{amsmath}
\usepackage{amsfonts}

\begin{document}
%
% paper title
% can use linebreaks \\ within to get better formatting as desired
\title{ContentFlow: Mapping Content to Flows in Software Defined Networks}

% author names and affiliations
% use a multiple column layout for up to two different
% affiliations

%\author{\IEEEauthorblockN{\hspace{2.5cm} Abhishek Chanda\IEEEauthorrefmark{1}\vspace{0.25cm}}
%\IEEEauthorblockA{\IEEEauthorrefmark{1}WINLAB\\ Rutgers University\\ New Brunswick, NJ 08901\\achanda@winlab.rutgers.edu} \and \IEEEauthorblockN{Cedric Westphal\IEEEauthorrefmark{2}\IEEEauthorrefmark{3}\vspace{0.25cm}}
%\IEEEauthorblockA{\IEEEauthorrefmark{2}Innovation Center\\ Huawei\\ Santa Clara, CA 95050\\ cedric.westphal@huawei.com} \and \IEEEauthorblockN{}
%\IEEEauthorblockN{Dipankar Raychaudhuri\IEEEauthorrefmark{1} \hspace{2.5cm}\vspace{0.25cm}}
 %\IEEEauthorblockA{\IEEEauthorrefmark{3}Dept of Computer Engineering\\ University of California, Santa Cruz\\ Santa Cruz, CA 95064\\ cedric@soe.ucsc.edu} }
\author{
\IEEEauthorblockN{Abhishek Chanda}
\IEEEauthorblockA{WINLAB, Rutgers University\\
North Brunswick, NJ, USA\\
achanda@winlab.rutgers.edu}
\and %\hspace{25mm}
\IEEEauthorblockN{Cedric Westphal\IEEEauthorrefmark{2}\IEEEauthorrefmark{3}}
\IEEEauthorblockA{\IEEEauthorrefmark{2}Innovation Center\\
Huawei Technology\\
Santa Clara, CA, USA\\
cwestphal@huawei.com}
\and
\IEEEauthorblockN{}
\IEEEauthorblockA{\IEEEauthorrefmark{3}Dept of Computer Engineering\\ University of California, Santa Cruz\\ Santa Cruz, CA 95064\\ cedric@soe.ucsc.edu}
}

% make the title area
\maketitle

\begin{abstract}
Information-Centric Networks place content as the narrow waist of the network architecture. This allows to route based upon the content name, and not based upon the locations of the content consumer and producer. However, current Internet architecture does not support content routing at the network layer. We present ContentFlow, an Information-Centric network architecture which supports content routing by mapping the content name to an IP flow, and thus enables the use of OpenFlow switches to achieve content routing over a legacy IP architecture.

ContentFlow is viewed as an evolutionary step between the current IP networking architecture, and a full fledged ICN architecture. It supports content management, content caching and content routing at the network layer, while using a legacy OpenFlow infrastructure and a modified controller. In particular, ContentFlow is transparent from the point of view of the client and the server, and can be inserted in between with no modification at either end. We have implemented ContentFlow and describe our implementation choices as well as the overall architecture specification. We evaluate the performance of ContentFlow in our testbed.
\end{abstract}

\begin{IEEEkeywords}
SDN, content management, network abstractions, cache, storage
\end{IEEEkeywords}

% For peer review papers, you can put extra information on the cover
% page as needed:
% \ifCLASSOPTIONpeerreview
% \begin{center} \bfseries EDICS Category: 3-BBND \end{center}
% \fi
%
% For peerreview papers, this IEEEtran command inserts a page break and
% creates the second title. It will be ignored for other modes.
\IEEEpeerreviewmaketitle

\section{Introduction}
% no \IEEEPARstart

Many new Information-Centric Network (ICN) architectures have been proposed to enable next generation networks to route based upon content names~\cite{NDN,ANRConnect,Paul2008Cacheandforward,Pursuit}. The purpose is to take advantage of the multiple copies of a piece of content that are distributed throughout the network, and to dissociate the content from its network location. However, the forwarding elements in the network still rely on location (namely IP or MAC addresses) to identify the flows within the network.

One issue is that content is not isomorphous to flows, especially when it comes to network policy. Consider as an example two requests from a specific host to the same server, one for a video stream and one for an image object. Both these requests could be made on the same port, say port 80 for HTTP as is most of the Internet traffic. From the point of view of a switch, without Deep Packet Inspection (DPI), they will look identical. However, the requirements for the network to deliver proper service to these flows are very different, one with more stringent latency constraints than the other. Flows are not the proper granularity to differentiate traffic or to provide different sets of resource for both pieces of content.

Yet, flows are a convenient handle as they are what switches see. Consider the recent popularity of Software-Defined Networking (SDN) and OpenFlow~\cite{McKeown2008OpenFlow} in particular. This framework introduces a logically centralized software controller to manage the flow tables of the switches within a network: by using the flow primitive to configure the network, a controller can achieve significant benefits. A mechanism is needed to bridge the gap between the flow view as used in SDN and achieving policy and management at the content level, as envisioned by ICNs.

We present in this paper {\em ContentFlow}, a network architecture which leverages the principles of SDN to achieve the ICN goal of placing content at the center of the network: namely, with ContentFlow, a centralized controller in a domain will manage the content, resolve content to location, enable content-based routing and forwarding policies, manage content caching, and provide the extensibility of a software controller to create new content-based network mechanisms. 
In order to provision an information-centric network architecture over a legacy IP underlay, a mapping mechanism is necessary to identify flows based upon the content they carry. In this paper, we propose an extension to the current SDN model to perform this mapping and integrate, in addition to the switching elements, some caching elements into the OpenFlow framework. We also observe that there should be a centralized content management layer installed in a controller. We propose such a layer that uses HTTP header information to identify content, route it based on $name$ and map it back to TCP and IP semantics so that the whole system can operate on an underlying legacy network without any modification to either clients or servers. ContentFlow can be inserted transparently and independently by an operator over its own domain.

\begin{figure}[h]
\centering
\includegraphics[width=3in]{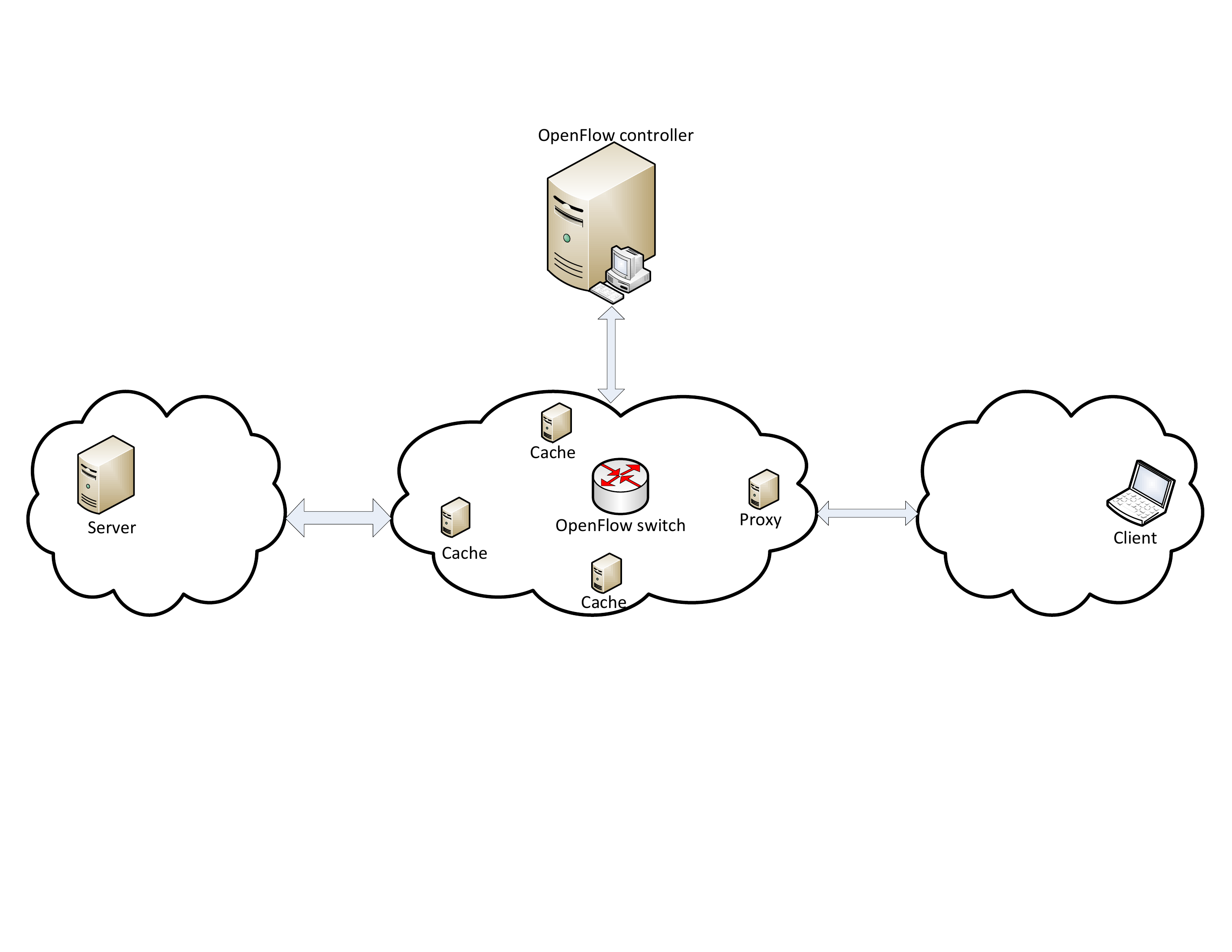}
\caption{End to end architecture}
\label{fig:system}
\end{figure}

To demonstrate the feasibility of our approach, we have implemented a content management framework on top of a SDN controller and switches which includes a {\em distributed, transparent caching mechanism}. Thus, the overall vision is to have a transparent caching network that can be placed between a service provider network and a consumer network as shown in figure \ref{fig:system}
 
Our contribution is thus to propose an architecture to bridge the gap between content and flows and enable the use of current SDN network elements in order to provide content-based routing, caching and policy management. We have implemented the architecture, focusing on HTTP traffic, and have demonstrated that it can operate transparently between unmodified clients and servers. We describe our implementation choices and present some evaluation results based upon the implementation.

The rest of the paper is organized as follows: Section~\ref{sec:related} examines some related work in this area. Section~\ref{sec:architecture} will show how to put these together to present an illustration of the capability enabled by our framework. Namely, we will construct a programmable content management module in the network controller in order to set up a transparent content caching network. Section~\ref{sec:management} describes the content management layer which forms the core of the whole architecture. We describe in Section~\ref{sec:eval} some details of our implementation and provide some evaluation results. Section~\ref{sec:cases} presents a walk through of some sample usage scenarios and explains how the architecture will handle those.  Section~\ref{sec:con} concludes.% the paper.

\section{Related Work}
\label{sec:related}

While information-centric networks and software defined networking has been explored as separate areas of research in the past, recent trend in adoption of OpenFlow has sparked research on combining the two. The benefits are very obvious: while SDN brings low maintenance and reduced network complexity, distributed caching brings low latency. Some recent findings in this vein includes \cite{Othman10} which proposes a application driven routing model on top of OpenFlow. Another notable work is \cite{Sakurauchi10} which proposes a system to dynamically redirect requests from a client based on content. Note that our approach is different from this, we have proposed a distributed caching system based on content based switching, thus extending \cite{Sakurauchi10}.

Software defined networking (SDN) decouples the control plane and the forwarding plane of a network. The forwarding plane then exports some simple APIs to the control plane, which then utilizes these APIs to provide desired network policy. The decoupling allows for the control plane to be written in software, and thus be programmable.

Current approaches for SDN, such as OpenFlow~\cite{McKeown2008OpenFlow}, focus on controlling switching element, and adopt a definition of the forwarding plane which takes {\em traffic flows} as the unit to which a policy is applied. 

However, this approach suffers from two limitations: 1) The network might include other non-forwarding elements; it is the common view of most future Internet architectures~\cite{NDN,Anand2011XIA,Paul2008Cacheandforward} and many commercial products~\cite{CiscoSRE} combine a switching element with additional capability, such as storage and processing; these capabilities need to be advertised to the control plane so that the programmable policy takes them into account. 2) The network might need to consider different policy for different objects at a finer granularity. Currently, OpenFlow ignores the specific content and provides a forwarding behavior that is per-flow based, where the flow is defined as a combination of the source and destination MAC and IP addresses and protocol ports plus some other fields such as MPLS tags, VLAN tags. Many architectures attempt to place content as the focus of the architecture~\cite{Jacobson2009Networking,Ghodsi2011InformationCentric,Koponen2007Dataoriented,Pursuit} and this is not supported by the current SDN proposals.

SDN has been considering L4-L7 extensions (say by Citrix or Qosmos), but those often take the form of DPI modules which do not integrate the routing decisions based upon the observed content. To operate on the granularity of content, many new architectures~\cite{NDN,Anand2011XIA,Paul2008Cacheandforward,Ghodsi2011InformationCentric,CBMEN} have been proposed, which make identifying content and forwarding based upon data possible. See~\cite{Ahlgren2012Survey} for a short survey on ICNs. However, these architectures typically require to replace the whole network while our architectural model is based upon inserting a content-management layer in the SDN controller.

Other proposals to extend the SDN framework include integration with the application layer transport optimization~\cite{Xie2012Use}. In an earlier paper~\cite{Chanda2012Content}, we have discussed extending the definition of network elements to include caches. However, this still requires a mapping layer between content and flow which we introduce here. \cite{Popa2010HTTP} proposed to use HTTP as the basis for an ICN, and our work finds inspiration in this idea, as we focus on HTTP content, in our implementation in particular. The extensions to HTTP from~\cite{Popa2010HTTP} could be used in ContentFlow, but would require modifications of the end points.

\section{Description of the architecture}
\label{sec:architecture}

\subsection{Overview}

The major design philosophy here is to implement content-based management using existing elements as much as possible. We present here the ContentFlow architecture, which enables allocation of content-dependent flow headers. We use a combination of the source, destination ports and IPv4 addresses, but not that we could also use IPv6 addresses and use the lower bits in the IPv6 address to embed content information.
 
The key design concern is to allow multiple flows from the same source/destination pair and to provide this handle to the different network elements. We consider both switches and caches to be network elements. Thus we need to identify a piece of content, assign it a flow identifier for routing purpose, yet at the same time, provide a caching element with this flow identifier and the original content name, so as to demultiplex content in the cache.

\begin{figure}[h]
\centering
\includegraphics[width=3in]{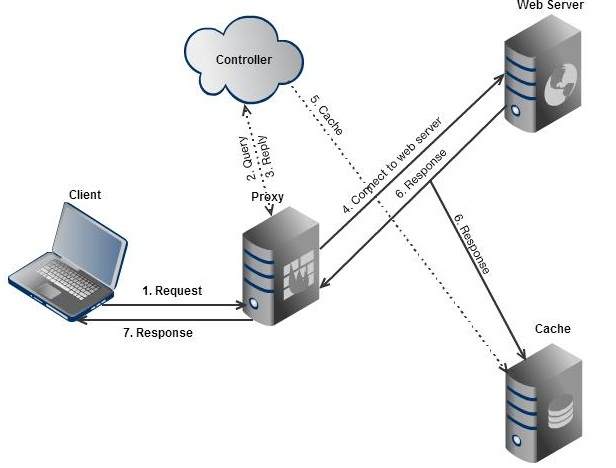}
\caption{Overview of the architecture}
\label{fig:arch}
\end{figure}

We perform a \emph{separation of content and its metadata} as early as possible. Thus, we try to separate out the metadata for the content close to the source and put it in the control plane. In our implementation, the metadata consists of the file name parsed from the HTTP GET request and the TCP flow information. Thus, it is a five tuple of the form \emph{$\langle file~ name, destination ~IP, destination~ port, source~ IP, \\source~ port \rangle$}. A high level overview of the architecture is shown in figure \ref{fig:arch}.

One key aspect is that content routing in a SDN network requires to proxy all TCP connections at the ingress switch in the network. This is due to the TCP semantics: a session is first established between a client and the server, before an HTTP GET is issued to request a specific content. Routing is thus performed \emph{before} the content is requested, which prohibits content routing. On the other hand, content-routing requires {\em late binding} of the content with the location. Proxying TCP at the ingress switch ensures that no TCP session is established beyond the ingress switch. Only when the content is requested, then the proper route (and TCP session) can then be set-up.

Figure~\ref{fig:arch} shows the different architecture elements of ContentFlow: a proxy at the ingress to terminate the TCP sessions and support late binding; OpenFlow-enabled switching elements; caching and storage elements which can transparently cache content; a content controller expanding an OpenFlow controller to include content features; and the protocols for the controller to communicate with both the caches and the proxies (the controller-to-switch protocol is out-of-the-box OpenFlow). The different elements and protocols are detailed below. 

\begin{figure*}
\centering
\includegraphics[width=7in]{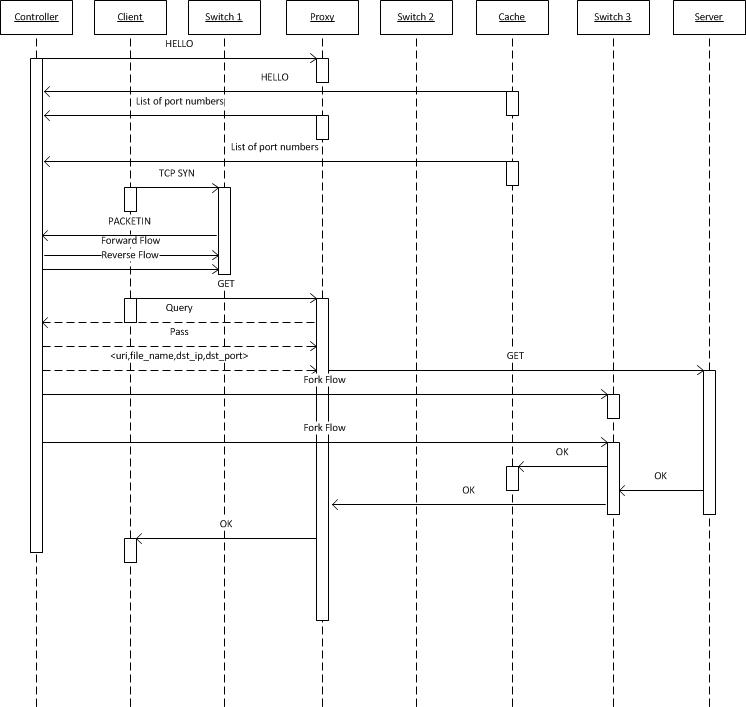}
\caption{Sequence diagram of the system}
\label{fig:sequence}
\end{figure*}

The system works as follows. A request for content entering the network for the first time will be a cache miss. Thus, the request will be redirected to the original destination server. Once the content has been cached, subsequent requests will be redirected to the cache if the controller decides so. 

\begin{enumerate}
\item When a client (in the client network) tries to connect to a server (in the provider network), the packet reaches the ingress switch of the ContentFlow network. This switch does not find a matching flow and sends this packet to the ContentFlow controller.
\item The controller installs static rules into the switch directly connected to the client to forward all HTTP traffic from the client to the proxy (on the selected port) and vice-versa. The TCP proxy terminates all TCP connections from the client. At this point, the client sets up a TCP session with the proxy and believes that it has a TCP session with the server.
\item The client issues an HTTP request which goes to the proxy.
\item The proxy parses the request to extract the name of the content and the web server to which the request was sent.
\item The proxy queries the controller with the file name (as a URI) asking if the file is cached somewhere in the network. The query also include the source of the request and the destination.
\item The controller then determines whether the content from the web server should (or should not) be cached and which cache to use. To redirect the content to the selected cache, it computes a forking point (forking switch) where the connection from the web server to the proxy should be duplicated towards the cache as well.
\item The controller installs a rule in the forking switch and invokes the cache. The controller notifies the cache of the content name and of the flow information to map the content stream to the proper content name. The controller also records the location of the cached content.
\item Since the content is not already cached, the controller returns "none" to the proxy. The controller gives the flow information to use to the proxy (destination port number). The proxy directs the connection to the original web server.
\item The cache saves the content with the file name obtained from the controller. Upon receiving the whole content, it ACKs to the controller, which updates its content location state.
\item The other copy of the content goes back to the proxy and in the egress
switch, it hits the reverse flow which re-writes its destination IP and port to that of the requester.
\end{enumerate}
For the cache hit case, steps $1$ to $5$ are the same and the controller returns a cache IP in step $5$. The proxy connects to the cache which serves back the content. On its way back, the content hits the reverse flow and the source information is re-written making the client believe that the content came from the original server.
Figure \ref{fig:sequence} shows how the system works.

\begin{figure}[h]
\centering
\includegraphics[width=3in]{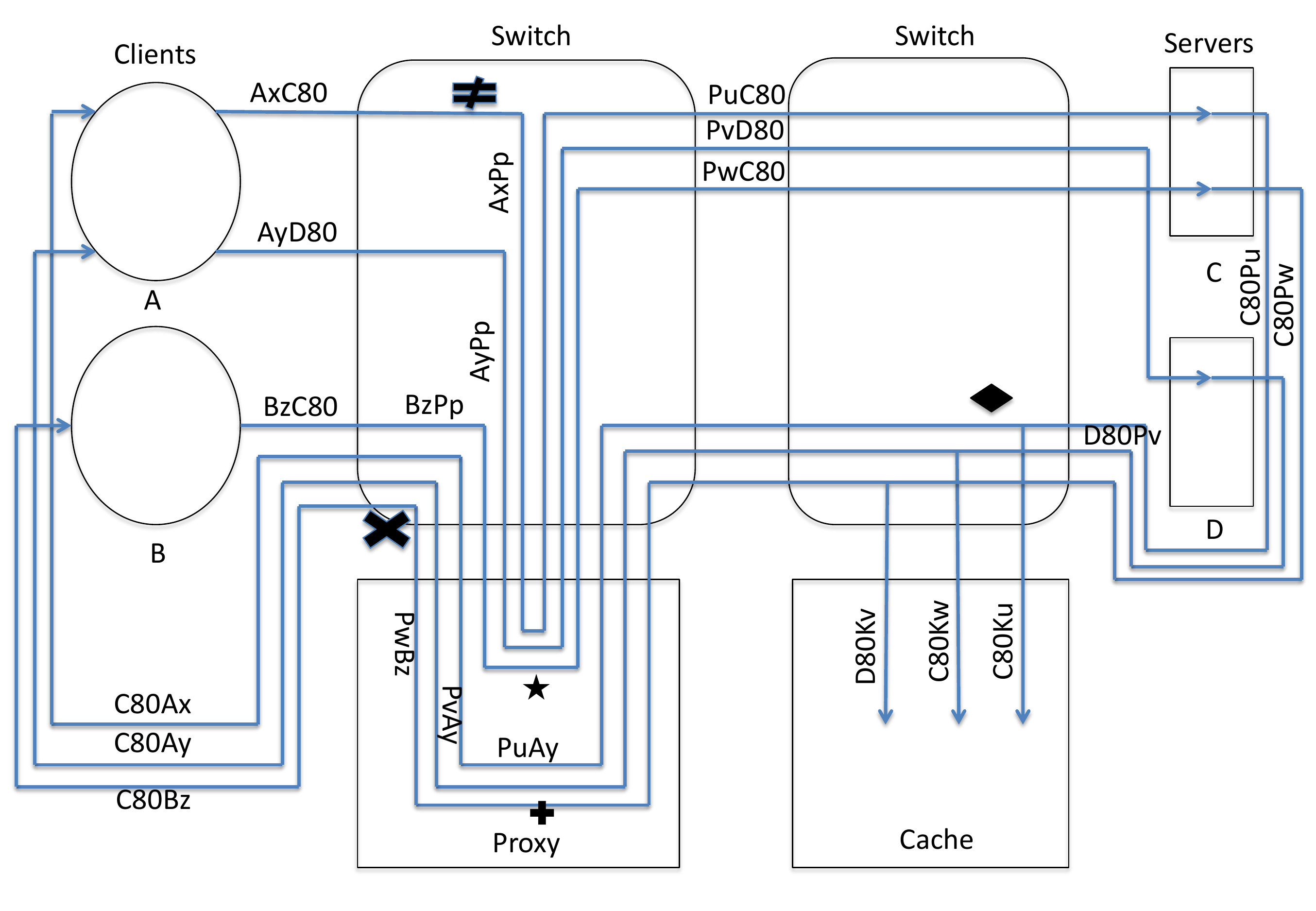}
\caption{Mapping of Content to Flows in ContentFlow: Two Clients A and B attempt to fetch 3 pieces of content; A fetches one each from Server C and D, and B gets one from C. The flow goes from source IP A, port x, to destination IP C, port 80. This is labelled as AxC80 on the figure. The path of the requests follow the thin arrow going through the ingress switch, the proxy, etc., to the server and back. Rules are applied along the path: ($\neq$) static rule to forward packets on port 80 from the client to the proxy. AxC80 becomes AxPp; ($\star$) Proxy terminates the TCP connection; Starts a TCP connection on src port u (assigned by the controller) to server C port 80, flow becomes PuC80; ($\lozenge$) catches return flows at the forking switch set by the controller, which duplicates packets towards cache K; (\textbf{+}) is the reciprocate of ($\star$) and forwards packets back to the client; (\textbf{x}) returns the server address in the source field. This is set by the controller to map the src port from proxy to switch back to the original server/port combination.}
\label{fig:ifip-diagram}
\end{figure}

\subsection{Proxy}

The proxy is a transparent TCP proxy that is located in the caching network. In our setup, OpenFlow is used to make the proxy transparent by writing reactive flows to appropriate switches.
%When the switch (on which the client is connected) connects to the controller, the controller writes a set of static rules to re-direct all web traffic from the client to the proxy and reverse.
%Here, the switch essentially works as a NAT which does both SNAT and DNAT.
The proxy is the primary device that is responsible for separating out content metadata and putting it on the control plane, thus it must intercept packets from the client and parse relevant information. The proxy communicates with the controller through REST API calls (in the future, this could be integrated in the OpenFlow protocol). Once the client has setup a connection with the proxy, it issues a HTTP request which the proxy intercepts. The proxy parses the request to extract content metadata and queries the controller. Algorithm \ref{alg:proxy} describes how the proxy works.

While booting up, the proxy starts listening on a set of ports. It also reads a configuration file to know the IP address and port of the OpenFlow controller to which it connects. The proxy then establishes a session with the controller and gives it a range of usable port numbers to be used as content handles. Now, when a client tries to connect to a server, the proxy parses the requests and forward the requests for content to the controller. The controller picks the first unused port number and redirects the connection to that port. From this point, this port number acts as a handle for the content name. The controller maintains a global mapping of the form $\langle content\_name, client\_ip\_port, server\_ip\_port, handle\rangle$ which can be queried either using $content\_name$ or $handle$ to retrieve the other parameters.

\begin{algorithm}
    \SetAlgoLined
    Listen on proxy port;\\
    \eIf{a GET request arrives} {
        Parse the file name from the request;\\
        Construct content name by combining the destination URI and the file name;\\
        Query controller with the content name;\\
        \eIf{the controller returns an IP address and port} {
            Redirect all requests to that IP address/port combo;
        }
        {
            Update controller with the file name;\\
            Pass the request unmodified;
        }
        Add the port to the list of free ports;\\
        Update the controller;\\
        }
    {
    Do not proxy;
    }
 \caption{Proxy Algorithm}
 \label{alg:proxy}
\end{algorithm}

%\vspace{-5pt}

\subsection{Cache}
In our design, when the cache receives a content stream to store, the cache will see only responses from the web server. The client's (more accurately, proxy's) side of the TCP session is not redirected to the cache. This scenario is not like generic HTTP caching systems like Squid which can see both the request and response and cache the response. In our design, we want to avoid the extra round trip delay of a cache miss, so we implemented a custom cache that can accept responses and store them against request metadata obtained from the controller. The cache implements algorithm \ref{alg:cache}.

We handle streaming video as a special case which is delivered over a HTTP 206 response indicating partial content. Often, in this case, the client will close the connection for each of the chunks. Thus, when the cache sees a HTTP 206 response, it parses the \texttt{Content-range} header to find out how much data is currently being transmitted and when that reaches the file size, it knows it has received the complete file and then it can save it.

\begin{algorithm}
    \SetAlgoLined
    Listen on cache port after receiving message from controller with (filename, server IP, dest port) filter;\\
    Start webserver on cache directory;\\
    \eIf{a HTTP response arrives} {
        \eIf{The response is a 206 partial content} {
         Extract the content-range header to know current range being sent;\\
         \While{Server has data to send} {Append to a file}
        \eIf {the controller has sent a file name matching the flow filter} {
            Save the response with the file name;
        }
        {
            Discard the response\;
        }
        } {
        Lookup the source IP of the response;\\
        \eIf {the controller has sent a file name matching the flow filter} {
            Save the response with the file name;
        }
        {
            Discard the response\;
        }
    } }
    {
        Serve back the file using the webserver\;
    }
    \caption{Cache Algorithm}
        \label{alg:cache}
\end{algorithm}

%\vspace{-5pt}

\subsection{Controller}

The controller can run on any device that can communicate with the switches and in our case, the caches (in most cases, it is placed in the same subnet where the switches are). It maintains two dictionaries that can be looked up in constant time. The $cacheDictionary$ maps file names to cache IP where the file is stored, this acts as a global dictionary for all content in a network. $requestDictionary$ maps destination server IP and ports to file name, this is necessary to set up flow rules and forward content metadata to the cache when it will save a piece of content. The controller algorithm is described in \ref{alg:controller}. The next section describes the content management layer that runs in the controller.\\

\begin{algorithm}
    \SetAlgoLined
    Install static flows for proxy forwarding in the switch to which the client is connected;\\
    $cacheDictionary \gets \{ \}$\\
    $requestDictionary \gets \{ \}$\\
    \If{proxy queries with a file name} {Lookup cache IP from $cacheDictionary$ and send back cache IP}
    \ElseIf{proxy sends content meta data} {Insert file name and destination IP to $requestDictionary$\\
                                            Select cache based upon caching policy and availability\\
                                            Compute the forking point for a flow from destination IP to proxy and destination IP to cache\\
                                            Push flows to all switches as necessary\\
                                            Invoke the cache and send it the file name from $requestDictionary$\\
                                            Insert the file name and cache IP in $cacheDictionary$ }
    \caption{Controller Algorithm}
    \label{alg:controller}
\end{algorithm}

%\vspace{-5pt}
\section{Content management layer and ContentFlow controller}
\label{sec:management}

As mentioned earlier, we propose an augmentation to a standard OpenFlow controller layer to include content management functionality. This layer handles the following functionality:
\begin{itemize}
\item \textbf{Content identification:} we propose content identification using HTTP semantics. This indicates, if a client in a network sends out a HTTP GET request to another device and receives a HTTP response, we will conclude that the initial request was a content request which was satisfied by the content carried over HTTP (however, the response might be an error. In that case we will ignore the request and the response). Further, we propose that this functionality should be handled in the proxy since it is directly responsible for connection management close to the client. The content management layer gathers content information from the proxy which parses HTTP header to identify content.
\item \textbf{Content naming:} as with a number of content centric network proposals, we propose that content should be named using its location. Thus, if an image of the name $picture.jpg$ resides in a server whose name is $www.server.com$ in a directory called $pictures$, the full name for the specific content will be $www.server.com/pictures/picture.jpg$. Such a naming scheme has several advantages; 1) it is unique by definition, the server file system will not allow multiple objects of the same name to exist in a directory. Therefore, this naming scheme will allow us to identify content uniquely. 2) this scheme is easy to handle and parse using software since it is well structured. 3) this scheme is native to HTTP. Thus it allows us to handle HTTP content seamlessly. As mentioned before, in our implementation, the content name is derived by parsing HTTP header for requests.
\item \textbf{Manage content name to TCP/IP mapping:} to enable the content management mechanism to work on a base TCP/IP system, we need to map content semantics to TCP/IP semantics (end point information like port and IP) and back. The content management layer handles this by assigning a port number to a specific content; the data structure can be looked up using the port number and server address to identify the content. The OpenFlow flows can be written to relevant switches based on that information. The whole process is described in figure \ref{fig:ifip-diagram}. Note that, the number of (server address,port combinations) in the proxy is finite and might be depleted if there are too many requests. Thus, freed up port numbers should be reused and also, the network admin should use sufficient number of proxies so that the probability of collision is sufficiently small.
\item \textbf{Manage Content Caching Policy:} the $cacheDictionary$ can be expanded to achieve the desired caching policy, but taking all the distributed caching available into account. For instance, the controller can append to each content some popularity information gathered from the number of requests, and thus decide which content to cache where based upon user's request history. 
\end{itemize}

It is easy to note here that, since the number of ports in a proxy is finite, there is a non zero probability that a large number of requests to a specific server might deplete the pool of possible (port,server) combinations. 
\begin{itemize}
\item \textbf{Ensuring availability:} since there is a limit on the number of potential requests to a given server, we must make sure there are enough port numbers to accommodate all possible content in our network. Since contents map to a port/proxy/server combination, the number of concurrent connections to a server is as limited by the number of available src port between a given proxy/server pair. Further, the number of proxies can be increased, by adding virtual proxies with different addresses, and thus increasing the flow space. IPv6 could be used there as well to increase the flow space size and demultiplex different content over different flows. The use of a flow filter for a restricted period of time can be translated as a queueing theoretic issue and has been proposed in the case of network address translation for IPv4-IPv6 migration in~\cite{Westphal2010Queueing}. The analysis therein translates directly into dimensioning the number of concurrent content requests that a proxy/server pair can perform.
\item \textbf{Ensuring consistency:} the consistency constraint implies all clients should receive their requested content only and not each other's. In our system, this translates to the constraint that once a port number is designated for a content, it should not be re-used while that content is live in the network. This is ensured using the controller. The controller maintains a pool of port numbers from which it assigns to content. Once a port number is in use, it is removed from the pool. When the client closes the connection to the proxy, the proxy sends a message to the controller and updates the list of free port numbers, which the controller puts back in the pool.
\end{itemize}
Thus while the proposed system ensures availability and consistency, it is not fully partition tolerant since the whole architecture is sensitive to the latency between the OpenFlow controller and the network elements. This limitation however, is from OpenFlow and not our architecture which leverages OpenFlow.

\section{Implementation and Evaluation}
\label{sec:eval}

\begin{figure*}
\centering
\includegraphics[scale=0.5]{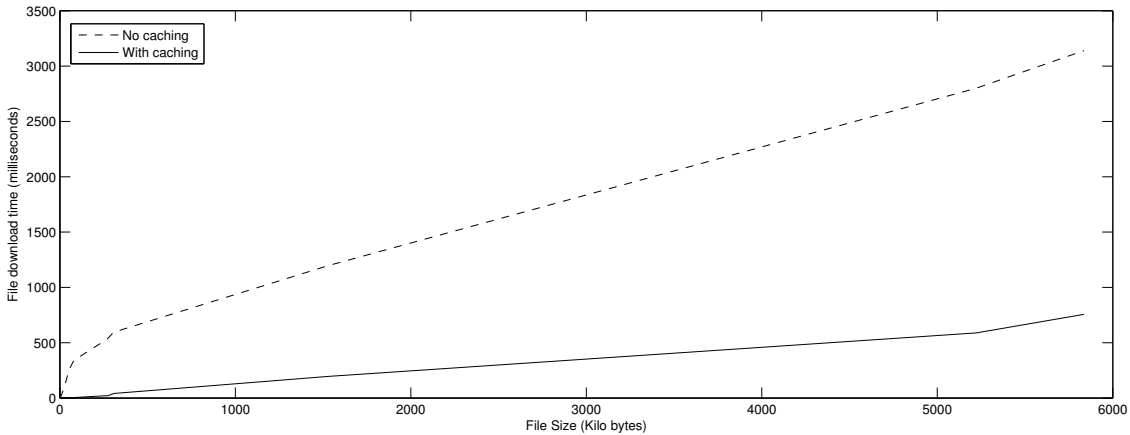}
\caption{Variation of access delay (as seen on the client) with file size of content}
\end{figure*}

To evaluate the merits of ContentFlow, we implemented a basic use case of in-network storage virtualization. This is a simple application of the content-flow mapping. We hope other applications can take advantage of the framework. We implemented our proposed architecture on a small testbed. The testbed has a blade server (we would call it \emph{H}) running Ubuntu. The server runs an instance of Open vSwitch which enables it to act as an OpenFlow switch. \emph{H} runs three VMs each of which hosts the cache, the proxy and the client and also the FloodLight controller. This setup is placed in Santa Clara, CA.

%To evaluate the merits of in network storage virtualization, we implemented our proposed architecture on a small testbed. The testbed has two blade servers (we would call them \emph{H1} and \emph{H2}) running Ubuntu. Each of the server hosts multiple virtual machines, which run each of the components of the architecture. Each blade server has two ethernet devices, \emph{eth1} of the servers are directly connected to each other and forms the data plane for the architecture. \emph{eth0} of both have public IP addresses and are connected to the top of the rack switch. This forms the control plane for the architecture. Both \emph{H1} and \emph{H2} run instances of Open vSwitch which enables them to act as OpenFlow switches. On each host, \emph{eth1} is configured as a trunk. \emph{H1} runs a VM which hosts the web server, it also runs FloodLight (the OpenFlow controller). \emph{H2} runs three VMs each of which hosts the cache, the proxy and the client
%\begin{figure}[H]
%\centering
%\includegraphics[scale=0.5]{topology.jpg}
%\end{figure}

The major components of the system are described in the following subsections.
%\begin{itemize}
\subsection{Content management layer}
We used FloodLight as the OpenFlow controller. FloodLight allows loading custom modules on the controller platform which can then write flows to all connected switches. We implemented the content management layer as a  module to do content based forwarding on top of FloodLight. It subscribes to \emph{PACKET\_IN} events and maintains two data structures for lookup. The \emph{requestDictionary} is used to map $\langle client,server \rangle$ pairs to request file names. This data structure can be queried using REST API to retrieve the file name corresponding to a request. The \emph{cacheDictionary} holds mapping of content and its location as the IP and port number of a cache.
\subsection{Proxy}
The proxy is written in pure Python and uses the \emph{tproxy} library. The library provides methods to work with HTTP headers, note that there is no way to access any TCP or IP information in the proxy. The proxy uses the controller's REST API to communicate with it. It should be instantiated with the command
\begin{verbatim}
sudo tproxy <script.py> -b
0.0.0.0:<port number>
\end{verbatim}
According to our implementation, the box running the proxy runs multiple instances of it on different ports. Each of those instances will proxy one $\langle client,server \rangle$ pair.
\subsection{Cache}
Our implementation of a cache is distinct from existing Internet caches in a number of ways: it can interface with an OpenFlow controller (running the content management module), on the flip side this does not implement usual caching protocols simply because it does not need to. Standard Internet caches see the request and if there is a miss, forwards it to the destination server. When the server sends back a response, it saves a copy and indexes it by the request metadata. Thus, they can setup a TCP connection with then server and use the socket interface to communicate. In our case, the cache sees only the response and not the request. Since it always get to hear only one side of the connection, it cannot have a TCP session with the server and so, cannot operate with a socket level abstraction. Thus, the cache must listen to a network interface and read packets from it. The cache has a number of distinct components as described below:
\begin{itemize}
\item \textbf{The Redis queue}
There is a Redis server running in the backend which serves as a simple queueing mechanism. This is necessary to pass data (IP addresses) between the grabber module and the watchdog module. The grabber module (described below) can put IP addresses in the queue which can be read from the watchdog module.
\item \textbf{The grabber module}
The grabber module is responsible for listening to an interface and to read (and assemble) packets. It is written in C++ and uses the \emph{libpcap} library. The executable takes an interface name as a command line argument and starts to listen on that interface. It collects packets with the same ACK numbers, when it sees a FIN packet, it extracts the ACK number and assembles all packets which has that ACK number. In this step, it discards duplicates. Note that, since there is no TCP connection, the cache won't know if some packets are missing. It then extracts data from all those packets and writes back to a file in the disk with a default name. It also puts the sender's IP in the Redis queue.
\item \textbf{The watchdog module}
This module communicates with the controller using a set of REST calls. It is written in Python and uses the \emph{inotify} library to listen on the cache directory for file write events. When the grabber module writes a file to the disk, the watchdog module is invoked. It calls the controller API to get the file name (using the IP from the Redis queue as parameter), it then strips all HTTP header from the file, changes its name and writes it back. After the file is saved, it sends back an ACK to the controller indicating that the file is cached.
\item \textbf{The cache server module}
This module serves back content when a client requests. This is written in Python and is an extended version of \emph{SimpleHTTPServer}.
\end{itemize}
%The cache is written in C++. It listens on an ethernet device on the VM and collects packets. When it sees a TCP FIN flag for a connection, it assembles data from the connection. It discards retransmitted packets and then re-arranges packets based on sequence number and writes to a file on the disk. A python script listens for file write events on the disk and when the cache saves the file, it gets triggered. It gets the file name from the controller and saves the file in the disk.
%\end{itemize}

%Figure 2 shows the setup in more detail.
%\begin{figure}[h]
%\centering
%\includegraphics[scale=0.5]{topo_2.jpg}
%\end{figure}
We placed $12$ files of different sizes, from $2Kb$ to $6Mb$ on a web server located in New Brunswick, NJ. These files are then accessed from our client in Santa Clara, CA which opens up a regular browser and access the files over HTTP. We turned off browser cache for our experiments to ensure that the overall effect is only dues to our cache. FireBug is used to measure content access delay in the two cases case, once when there is a cache miss (and the content gets stored in the cache) and the cache hit case (when the content is delivered from the cache). Caching content (and its performance benefit) is well known and we do not claim that our method innovate in this dimension. We only use this to demonstrate that our architecture works in offering content APIs to a controller and that we have successfully extended the SDN architecture to support content-based management and information-centric network ability.

We can also do a back of the envelope delay analysis given Figure~\ref{fig:sequence}. We compare and contrast three cases, in each case $TCP(A,B)$ represents the delay for establishing a TCP session between $A$ and $B$ and later to tear it down, $F(A,B)$ is the delay to download a file from $A$ to $B$, $Delay(Proxy)$ is the processing delay at the proxy.
\begin{itemize}
\item Case 1 is when a client accesses a file directly without a cache or a proxy. In this case, the total delay is approximately equal to $TCP(Client,Server)+F(Server,Client)$.
\item Case 2 is when the client uses a proxy (and an OpenFlow controller) to connect to the server. In this case, total delay is $TCP(Client,Proxy)+Delay(Proxy)+TCP(Proxy,Server)+F(Server,Proxy) + F(Proxy,Client)$
\item Case 3 is our design with a proxy and a cache. Here delay $TCP(Client,Proxy)+Delay(Proxy)+TCP(Proxy,Cache)+F(Server,Proxy) + F(Proxy,Client)$
\end{itemize}
From the expressions, we see that case 1 has the highest delay followed by case 2 and case 3, assuming $Delay(Proxy)$ is negligible and that the proxy and cache is located in the same network as the client which makes file transfer delay between them very small. We noticed that, when content size is greater than $5kb$, all the expressions are dominated by $F(A,B)$ and $Delay(proxy)$ can be ignored. However, if the proxy is overloaded, this will increase resulting in case 2 and 3 having higher delay than case 1, which is not desireable. Two possible ways to reduce computation at the proxy is to have static rules for popular content at the switch. Or a bloom filter having a list of content cached in the network can be placed at the proxy to avoid controller lookup (with a default port number to the cache). This method introduces another level of indirection as a cache on the proxy.

\section{Discussion}
\label{sec:cases}

As we mentioned before, when a content request enters the network, the network assigns a name and port number in anticipation. When the actual content enters the network, it is demultiplexed using the name and port number. Thus, the cases of multiple servers, clients or content can be converted to simpler cases of single devices (real or virtual) communicating.

In this section we present some sample usage scenarios of the proposed architecture.
\begin{itemize}
\item The simplest scenario is that of one client and one server where the on the client only one process communicates with the server. Given the proposed architecture, this case is trivial. One unused proxy port will be assigned to the client server pair and the content will be named as described before.
\item The case of multiple clients and a single server is also simple. Each of the client server pairs will be assigned a port number which will be used a demux handle throughout the network, along with the content name.
\item A more complex scenario is that of multiple processes on a client, each talking to the same server. More often than not, they will request multiple content and will run on different ports. We argue that our architecture will handle this scenario seamlessly since it identifies \emph{consumer} and \emph{producer} in the ICN sense by a combination of IP address and port number. Thus, it will see a number of virtual clients, each with the same IP address trying to connect to a server. Each of the virtual client -server pairs will be assigned a port number and their content will be named.
\item A related scenario is that of multiple files from the server to a single client. This case can be treated as multiple virtual servers, each communicating a piece of content to a client.
\item Finally, the case of multiple clients and multiple servers. It is easy to see that this case is a combinatorial extension of the previous cases and will be handled similarly.
\end{itemize}

\section{Conclusion and future directions}
\label{sec:con}

In this paper, we have proposed and evaluated ContentFlow, a generalization of the SDN philosophy to work on the granularity of content rather than flow. This hybrid approach enables the end user to leverage the flexibility a traditional SDN provides coupled with the content management properties an ICN provides. We have implemented ContentFlow in a small scale testbed, and demonstrated that it is able to perform network-layer task at the content level. 

Some immediate questions point to directions for future work: how can we scale ContentFlow to handle large traffic loads? Can we improve the proxy hardware and can we distribute the content management functionality of the controller in order to reduce the latency?  How does the caching policy impact the performance as observed by the end user? And can the set of primitives be expanded further beyond flows and contents, to include for instance some more complex workloads requiring synchronization of a wide set of network resources, including storage, compute and network? We plan to handle these problems as a natural extension to this work.

% conference papers do not normally have an appendix

% trigger a \newpage just before the given reference
% number - used to balance the columns on the last page
% adjust value as needed - may need to be readjusted if
% the document is modified later
%\IEEEtriggeratref{8}
% The "triggered" command can be changed if desired:
%\IEEEtriggercmd{\enlargethispage{-5in}}

% references section

% can use a bibliography generated by BibTeX as a .bbl file
% BibTeX documentation can be easily obtained at:
% http://www.ctan.org/tex-archive/biblio/bibtex/contrib/doc/
% The IEEEtran BibTeX style support page is at:
% http://www.michaelshell.org/tex/ieeetran/bibtex/
%\bibliographystyle{IEEEtran}
% argument is your BibTeX string definitions and bibliography database(s)
%\bibliography{IEEEabrv,../bib/paper}
%
% <OR> manually copy in the resultant .bbl file
% set second argument of \begin to the number of references
% (used to reserve space for the reference number labels box)

\bibliographystyle{ieeetr}
\bibliography{ContentFlow}

\begin{thebibliography}{10}

\bibitem{NDN}
``Named data networking.'' http://named-data.org/, Aug. 2010.

\bibitem{ANRConnect}
``{ANR Connect}, "content-oriented networking: a new experience for content
  transfer".'' http://www.anr-connect.org/, Jan. 2011.

\bibitem{Paul2008Cacheandforward}
S.~Paul, R.~Yates, D.~Raychaudhuri, and J.~Kurose, ``The cache-and-forward
  network architecture for efficient mobile content delivery services in the
  future internet,'' in {\em Innovations in NGN: Future Network and Services,
  2008. K-INGN 2008. First ITU-T Kaleidoscope Academic Conference},
  pp.~367--374, IEEE, May 2008.

\bibitem{Pursuit}
``{PURSUIT}: Pursuing a pub/sub internet.'' http://www.fp7-pursuit.eu/, Sept.
  2010.

\bibitem{McKeown2008OpenFlow}
N.~McKeown, T.~Anderson, H.~Balakrishnan, G.~Parulkar, L.~Peterson, J.~Rexford,
  S.~Shenker, and J.~Turner, ``{OpenFlow}: enabling innovation in campus
  networks,'' {\em SIGCOMM Comput. Commun. Rev.}, vol.~38, pp.~69--74, Mar.
  2008.

\bibitem{Othman10}
O.~M.~M. Othman and K.~Okamura, ``Design and implementation of application
  based routing using openflow,'' in {\em Proceedings of the 5th International
  Conference on Future Internet Technologies}, CFI '10, (New York, NY, USA),
  pp.~60--67, ACM, 2010.

\bibitem{Sakurauchi10}
Y.~Sakurauchi, R.~McGeer, and H.~Takada, ``Open web: Seamless proxy
  interconnection at the switching layer,'' in {\em Networking and Computing
  (ICNC), 2010 First International Conference on}, pp.~285 --289, nov. 2010.

\bibitem{Anand2011XIA}
A.~Anand, F.~Dogar, D.~Han, B.~Li, H.~Lim, M.~Machado, W.~Wu, A.~Akella,
  D.~Andersen, J.~Byers, S.~Seshan, and P.~Steenkiste, ``{XIA}: An architecture
  for an evolvable and trustworhty internet,'' in {\em Carnegie Mellon
  University TR CMU-CS-11-100}, Jan. 2011.

\bibitem{CiscoSRE}
``Cisco service ready engine.'' http://www.cisco.com/en/US/products/
  ps10598/prod\_module\_series\_home.html, downloaded June 2012.

\bibitem{Jacobson2009Networking}
V.~Jacobson, D.~K. Smetters, J.~D. Thornton, M.~F. Plass, N.~H. Briggs, and
  R.~L. Braynard, ``Networking named content,'' in {\em Proceedings of the 5th
  international conference on Emerging networking experiments and
  technologies}, CoNEXT '09, (New York, NY, USA), pp.~1--12, ACM, 2009.

\bibitem{Ghodsi2011InformationCentric}
A.~Ghodsi, T.~Koponen, B.~Raghavan, S.~Shenker, A.~Singla, and J.~Wilcox,
  ``{Information-Centric} networking: Seeing the forest for the trees,'' in
  {\em the Tenth ACM Workshop on Hot Topics in Networks HotNets 2011}, Nov.
  2011.

\bibitem{Koponen2007Dataoriented}
T.~Koponen, M.~Chawla, B.~G. Chun, A.~Ermolinskiy, K.~H. Kim, S.~Shenker, and
  I.~Stoica, ``A data-oriented (and beyond) network architecture,'' in {\em
  Proceedings of the 2007 conference on Applications, technologies,
  architectures, and protocols for computer communications}, SIGCOMM '07, (New
  York, NY, USA), pp.~181--192, ACM, 2007.

\bibitem{CBMEN}
``{CBMEN}: Content-based mobile edge networking.'' Solicitation Number:
  {DARPA-BAA-11-51}, http://www.darpa.mil/ Our\_Work/STO/Programs/Content-
  Based\_Mobile\_Edge\_Networking\_ (CBMEN).aspx, May 2012.

\bibitem{Ahlgren2012Survey}
B.~Ahlgren, C.~Dannewitz, C.~Imbrenda, D.~Kutscher, and B.~Ohlman, ``A survey
  of information-centric networking,'' {\em Communications Magazine, IEEE},
  vol.~50, pp.~26--36, July 2012.

\bibitem{Xie2012Use}
H.~Xie, T.~Tsou, H.~Yin, and D.~Lopez, ``Use cases for {ALTO} with software
  defined networks,'' Tech. Rep. draft-xie-alto-sdn-use-cases-00.txt, IETF
  Secretariat, Fremont, CA, USA, 2012.

\bibitem{Chanda2012Content}
A.~Chanda and C.~Westphal, ``Content as a network primitive.''
  {arXiv:1212.3341}, http://arxiv.org/abs/1212.3341, Dec. 2012.

\bibitem{Popa2010HTTP}
L.~Popa, A.~Ghodsi, and I.~Stoica, ``{HTTP} as the narrow waist of the future
  internet,'' in {\em Proceedings of the Ninth ACM SIGCOMM HotNets Workshop},
  (New York, NY, USA), 2010.

\bibitem{Westphal2010Queueing}
C.~Westphal and C.~E. Perkins, ``A queueing theoretic analysis of source ip
  nat,'' in {\em Communications (ICC), 2010 IEEE International Conference on},
  pp.~1 --6, may 2010.

\end{thebibliography}

% that's all folks
\end{document}